# Spintronic logic: from transducers to logic gates and circuits

Christoph Adelmann[1], Florin Ciubotaru[1], Fanfan Meng[1,2], Sorin Cotofana[3], and Sebastien Couet[1]

[1]Imec, 3001 Leuven, Belgium, adelmann@imec.be
[2]KU Leuven, Department of Electrical Engineering, 3001 Leuven, Belgium
[3]Delft University of Technology, Department of Quantum and Computer Engineering, 2628 CD Delft, The Netherlands,

**While magnetic solid-state memory has found commercial applications to date, magnetic logic has rather remained on a conceptual level so far. Here, we discuss open challenges of different spintronic logic approaches, which use magnetic excitations for computation. While different logic gate designs have been proposed and proof of concept experiments have been reported, no nontrivial operational spintronic circuit has been demonstrated due to many open challenges in spintronic circuit and system design. Furthermore, the integration of spintronic circuits in CMOS systems will require the usage of transducers between the electric (CMOS) and magnetic domains. We show that these transducers can limit the performance as well as the energy consumption of hybrid CMOS-spintronic systems. Hence, the optimization of transducer efficiency will be a major step towards competitive spintronic logic system.**

## I. Introduction

During the last decades, magnetic random-access memory (MRAM) has experienced rapid progress from fundamental science to integration as embedded memories in commercial systems-on-a-chip (SOCs). This success story has sparked intense interest to use magnetic degrees of freedom not only for information storage but also for information processing in spintronic logic circuits. Such circuits hold promise for high density combined with operation at ultralow power [1-3]. However, in contrast to the rapid advance of MRAM towards commercial applications, spintronic logic is today still at a rather conceptual level. As a matter of fact, no operating complex spintronic circuit has been reported so far and progress has been limited to the demonstration of simple proof of concept logic gates with comparatively large area and power consumption [4,5]. Moreover, different logic approaches using various magnetic excitations have been proposed and experimentally studied with a common framework for benchmarking and downselection only emerging [1,6].

To proceed from individual logic gates to ultimately competitive ultralow power circuits, a number of obstacles need to be overcome. The understanding of the obstacles can still be considered incomplete, and much research is still needed to fully recognize key roadblocks and devise possible solutions. Below, we list briefly review principal spintronic logic concepts and attempt to give an overview of the central challenges for operational and ultimately performant spintronic circuits.

## II. Circuit considerations for spintronic computation

Several approaches for spintronic computation have been proposed in the literature, based on domain walls [2,5,7], chirally-coupled nanomagnets [8], magnetic chiral solitons [9], spin waves [4,10,11], or magnetoelectric compounds [12]. Yet, all of these approaches share a number of challenges and pose similar questions in terms of system integration. In this section, we will introduce some of these generic challenges for spintronic circuits. In the next section, we will introduce transducer challenges that appear when spintronic circuits are integrated into a CMOS environment.

### A. *Logic gates, invert functionality, and functional scaling*

Today, CMOS circuits are based on complementary MOS transistors interconnected by metallic vias and wires. The scaling of the device density occurs via the miniaturization of transistors and interconnects, as well as via improvements of the layout of elemental logic gates. Both the miniaturization of transistors and interconnects face fundamental, technological, and economic challenges. Therefore, a long-term vision to continue circuit scaling and thus uphold Moore’s law has been, rather than scaling individual transistors, to replace logic gates or other building blocks with disruptive approaches that show a fundamentally different scaling behavior.

As an example, spintronic majority gates have been proposed as such building blocks for circuit synthesis and it has been argued that this may lead to circuit area and power reduction with respect to CMOS. The advantage of majority gates is that can be naturally implemented by magnetic devices, either using the thresholding behavior of nanomagnets [7,11] or the linear supposition of spin waves [10]. Other key logic functions, such as an inverter or an XOR gate are however difficult to implement in some approaches, for example domain-wall-based logic. Therefore, the ideal set of logic building blocks still needs to be determined. It can be envisaged that the set may depend on the technology and render certain approaches more favorable at a circuit level. It should be mentioned that there is no necessity to base the approach on majority gates and other (more expressive) logic gates can also become of interest.

### B. *Cascading, fan-out, and signal integrity*

To form circuits, logic gates need to be cascaded, providing both fan-out and ensuring signal integrity at the output. A minimum requirement is that the output signal of a logic gate can be transported in interconnects (see next section) and used as input signal in the next logic stage. Many logic gates however still employ electrical input and output signals using transducers. However, such approaches cannot be directly cascaded in the magnetic domain and require numerous transducers with potentially large energy, area, and delay overheads. By contrast, fan-out requires gain in the system, which can be difficult to obtain in all-magnetic systems. Therefore, additional elements such as amplifiers (*e.g.*, for spin waves) may become necessary for circuit operation, also to avoid signal decay throughput the circuit. Many spintronic computation approaches have not yet been worked out in sufficient detail to solve cascading and fan-out issues. Therefore, much additional work (both by simulations and experimental proof-of-concept demonstrations) are needed to better compare the advantages and disadvantages of different spintronic computation approaches.

### C. Interconnections

As shown below, a multitude of transducers between electric and magnetic domains can lead to large energy, area, and delay overheads and render spintronic circuits uncompetitive. Therefore, signals should propagate as much as possible in the magnetic domain, leading to the requirement of magnetic interconnects. While, *e.g.*, domain walls or spin waves can propagate in suitable magnetic conduits, propagation is comparatively slow. Moreover, multilevel magnetic interconnects, required for signal routing in more complex circuits are still an open issue. While such limitations can be mitigated by repetitions of certain parts of the circuits (rather than by the propagation of output signals), this leads to large penalties in terms circuit area and energy consumption. Hence, the conceptual development of complex magnetic signal routing and its benchmarking is an important topic that that needs to be resolved to enable competitive spintronic circuits.

## III. Transducers in hybrid spintronic-CMOS systems

Currently, there is no concept for a completely magnetic computer and therefore, spintronic circuits need to be interfaced with the SOC, which is based on electric signals. Hence transducers between electric and magnetic domains are required the interface between spintronic and electronic circuits. If signal propagation within a spintronic circuit is not possible in the magnetic domain only (*e.g.*, due to routing, fan-out, or signal propagation issues), transducers may also be employed within a spintronic circuit to enable charge-based signal routing between magnetic logic gates. A system level assessment of both domain-wall- as well as spin-wave-based hybrid CMOS-spintronic systems has found that in both cases, computation power/energy and delay are strongly determined by transducer performance [6,13]. While the intrinsic energy required to perform magnetic computation can be low, the total energy to generate the necessary magnetic excitations can still be large. As shown in Fig. 1, the power-delay product due to writing and reading logic signals with commercial MRAM cells is for many arithmetic circuits much larger than the computation in equivalent CMOS circuits [6]. This means that such spintronic circuits cannot compete with CMOS due to transducer overhead even if the spintronic computation itself requires negligible energy and delay. It shows that the development of spintronic

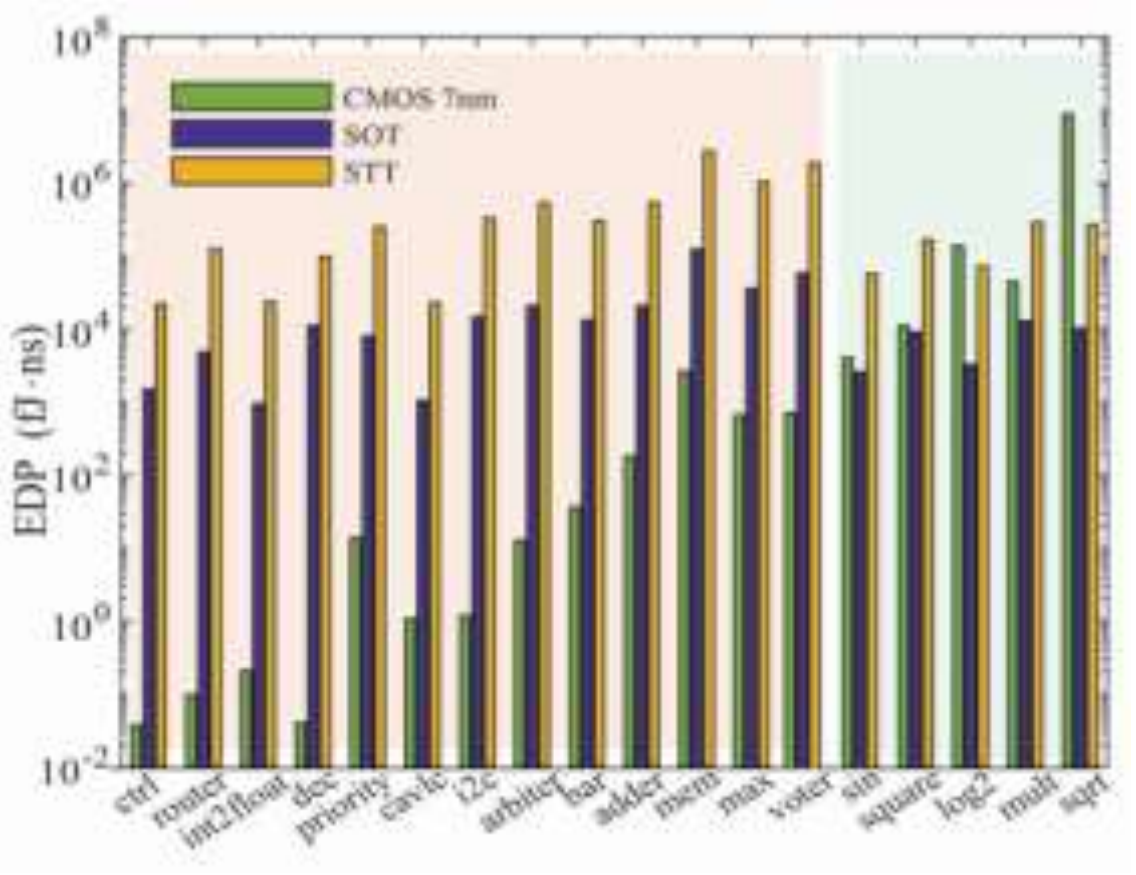

Fig. 1. Energy-delay product EDP of different arithmetic circuits implemented in a domain-wall-based spintronic logic technology, considering only input and output transducers. Already the transducer overhead EDP exceeds the EDP of the entire CMOS circuits. Domain wall propagation by currents in such devices further increases the EDP of the spintronic circuits by 1 to 2 orders of magnitude.

circuits needs to be complemented by the development of ultra-efficiency transducers.

The results in Fig. 1 suggest that the efficiency of current MRAM-derived transducers (writing by STT or SOT, reading by TMR) needs to be improved by several orders of magnitude before becoming competitive with CMOS. Magnetoelectric (capacitive) transducers promise the necessary efficiency but have not been developed to similar maturity. This will be a key issue for spintronic computation in general while also potentially providing interesting new routes for magnetic memory.